\documentclass[aps,pra,nofootinbib,twocolumn,superscriptaddress]{revtex4-1}
\usepackage{subfigure,graphicx} 
\usepackage{amsfonts,amssymb,amsmath}
\usepackage{flafter}
\usepackage{multirow}
\usepackage{txfonts}
\usepackage[unicode,breaklinks]{hyperref}
\usepackage{siunitx}
\newcommand{\papertitle}{Mitigating boundary effects in finite temperature simulations of false vacuum decay}
\hypersetup{
    unicode=true,
    plainpages=false,
    pdftitle={\papertitle},
    pdfauthor={Kate Brown, Ian G. Moss, Thomas P. Billam},
    pdfsubject={\papertitle},
    colorlinks=true,
    linkcolor=blue,
    citecolor=blue,
    filecolor=black,
    urlcolor=blue
}

\usepackage{ulem}

\newcommand{\be}{\begin{equation}}
\newcommand{\ee}{\end{equation}}
\newcommand{\bea}{\begin{eqnarray}}
\newcommand{\eea}{\end{eqnarray}}
\newcommand{\beal}{\begin{aligned}}
  \newcommand{\eeal}{\end{aligned}}

\begin{document} 

\title{\papertitle}
\author{Kate Brown}
\email{kbrown97123@outlook.com}
\affiliation{School of Mathematics, Statistics and Physics, 
Newcastle University, Newcastle upon Tyne, NE1 7RU, UK}

\author{Ian G. Moss}
\email{ian.moss@ncl.ac.uk}
\affiliation{School of Mathematics, Statistics and Physics, 
Newcastle University, Newcastle upon Tyne, NE1 7RU, UK}

\author{Thomas P.\ Billam}
\email{thomas.billam@ncl.ac.uk}
\affiliation{Joint Quantum Centre (JQC) Durham--Newcastle, School of Mathematics, Statistics and Physics, 
Newcastle University, Newcastle upon Tyne, NE1 7RU, UK}

\date{\today}

\begin{abstract}
The physics of false vacuum decay during first-order phase transitions in the early universe may be studied in the laboratory via cold-atom analogue simulators. However, a key difference between analogue experiments and the early universe is the trap potential confining the atoms. Rapid seeded bubble nucleation has been shown to occur at the boundary of typical trap potentials, obscuring the bulk bubble nucleation rate. This difficulty must be overcome in order to reliably probe the bulk bubble nucleation rate in an analogue simulator experiment. In this paper we show that, at finite temperature, this deleterious boundary nucleation can be mitigated by adding a `trench' to the potential, effectively screening the boundary with a region of higher atomic density. We show that this technique is effective in two different cold-atom analogue systems, but is not needed in ferromagnetic analogue simulators.
\end{abstract}

\maketitle

\section{Introduction}
\label{sec:intro}

The universe underwent many phase transitions as it evolved through the early thermal plasma of the hot big bang, with the nature of its earliest transitions still an open question \cite{Mazumdar:2018dfl,Hindmarsh2021}. Today, we may be able to observe signatures of a highly non-equilibrium history, in the form of cosmic strings \cite{Copeland:2009ga}, baryon asymmetry \cite{Caneletti:2024kww} and gravitational waves \cite{Hindmarsh:2013xza}. It is thus of interest to consider that some primordial phase transitions may have been first order, characterized by the universe supercooling into metastable ``false vacuum'' states and escaping via bubble nucleation.

The standard analytic treatment of bubble nucleation, established for first order phase transitions, and extended to vacuum decay in quantum field theory by Coleman et al. in the 1970s, seeks a semi-classical approximation to the field equations. The thermal and vacuum processes can be unified in the instanton approach, obtained by solving the relevant field equations in imaginary time~\cite{Coleman:1977py,Callan:1977pt}. The instanton approximation constitutes a valuable resource for predicting decay rates in quantum field theory, although it does not provide a real-time picture of the decay process, or address vacuum bubble correlations~\cite{Pirvu:2021roq}. Thus, the complementary approach of using an analogue system as a \textit{quantum simulator}~\cite{Altman:2021} to examine false vacuum decay in real time has gained traction in recent years~\cite{FialkoFate2015,FialkoUniverse2017,Braden:2017add,Braden:2018tky,Billam:2018pvp,Braden:2019vsw,Ng:2020pxk,Billam:2020xna,Billam:2021nbc,Billam:2021psh,Pirvu:2021roq,Braden:2022odm,Billam:2023,Jenkins:2023npg}. Ultracold atomic gases are a favored physical system with which to realize such analogues, owing to their ever-growing versatility and controllability, and were recently used to realize an analogue of thermally nucleated false vacuum decay in a ferromagnetic condensate~\cite{Zenesini:2024}. Theoretically, such systems are amenable to stochastic numerical modelling under the general umbrella of stochastic classical field methods~\cite{blakie_dynamics_2008}.

Unlike the early universe, cold-atom analogue systems are confined by traps, and so possess boundaries at which the atomic density falls to zero. Therefore, it is crucial to understand how false vacuum decay proceeds in the presence of these boundaries and to be able to create an analogue that reliably probes the bulk nucleation rate. While some previous works have investigated boundary effects in these analogues~\cite{Billam:2018pvp,Billam:2023}, to our knowledge the latter has not been achieved. The main result of this paper is to show that, in a quasi-two-dimensional setup, deleterious boundary effects can be suppressed by incorporating a \textit{trench} into a uniform `bucket'-type trapping potential~\cite{Gaunt2013} that screens the boundary of the trapping potential with a high-atomic-density region. Such a customised potential is likely realizable in quasi-two-dimensional geometries using digital micromirror device setups~\cite{Gauthier2016}. 

An extensively explored choice of cold-atom analogue is the pseudo-spin-1/2 setup of Fialko \textit{et al.}~\cite{FialkoFate2015,FialkoUniverse2017}. This approach makes use of two spin states of an optically trapped spinor condensate, with these components coupled using a time-modulated microwave field. When averaged over long timescales and expressed in terms of the relative phase between components, the resultant description possesses the textbook vacuum-decay landscape; a quasi-relativistic real scalar effective field exhibiting a metastable local minimum (false vacuum) and a stable global minimum (true vacuum) separated by a potential barrier. This effective description has been examined in depth in one-dimensional periodic systems at both zero and finite temperature~\cite{Braden:2017add,Braden:2018tky,Braden:2019vsw,Ng:2020pxk,Billam:2020xna,Billam:2021psh,Pirvu:2021roq,Braden:2022odm,Jenkins:2023npg}, with simulations reliably yielding bubbles in both regimes. However, the viability of this analogue hangs on the plausibility of overcoming a parametric instability present in the full time-dependent picture~\cite{Braden:2019vsw,Billam:2020xna,Billam:2021psh}. For this analogue, the potentially problematic seeding of rapid bubble nucleation at the trap boundary was noted in Ref.~\cite{Billam:2018pvp}.

A second analogue, also free from parametric instabilities, can be found in a spin-1 gas with three occupied Zeeman levels coupled via unmodulated Raman and radio-frequency mixing~\cite{Billam:2021nbc,Billam:2023}. In this analogue, one of the two real relative phases between the components serves as the effective scalar field degree of freedom exhibiting false and true vacua. This lesser-charted territory has thus far been explored for a one-dimensional zero-temperature system~\cite{Billam:2021nbc} and a two-dimensional finite-temperature system~\cite{Billam:2023}, the latter of which revealed the seeding of rapid bubble nucleation at the trap boundary. In particular, the inclusion of a simple square trapping potential in a quasi-two-dimensional (quasi-2D) gas, with parameters appropriate to a Lithium-7 system, was found to accelerate vacuum decay, with bubbles nucleating preferentially along the trap boundary to such an extent that a reliable measure of the bulk nucleation rate could not be isolated.

A third analogue, free from such parametric instabilities, and the only one that has been experimentally realized, uses a ferromagnetic condensate consisting of two hyperfine levels of sodium-23, with these components coupled by unmodulated microwave mixing~\cite{Zenesini:2024}. In this analogue, the relative population imbalance between the components is the effective field exhibiting false and true vacua. Notably, Ref.~\cite{Zenesini:2024} used a quasi-one-dimensional harmonic trapping potential; this setup has the advantage that boundary effects are not an issue since bubble nucleation is observed to primarily take place in the highest density region surrounding the trap minimum. If the ferromagnetic system is extended to two dimensions, with a bucket trap, one might predict that bubbles likewise avoid nucleating in the low density region beside the wall, in contrast to the other systems.

\begin{figure}[ht]
    \centering    
    \includegraphics[width=0.9\linewidth]{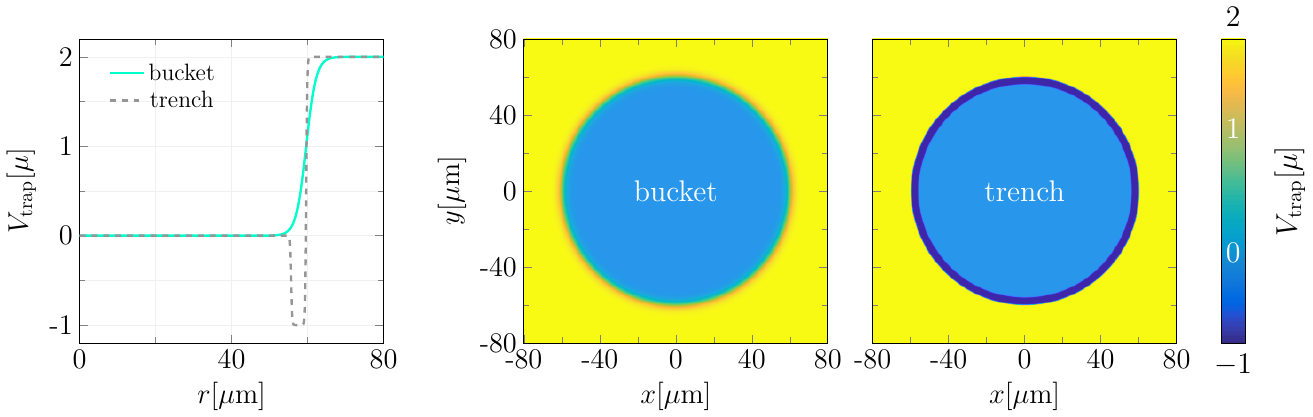}
    \caption{The circular trapping potential $V_{\text{trap}}$ given by Equation \eqref{eq:trap}, shown for the potassium-39 parameters listed in Table \ref{tab:K}. We examine two trap variations: a `bucket' trap, where $r_a = r_b$, and a `trench' trap, where $r_a < r_b$. Left: trapping potential as a function of radius. Right: The full spatial profile of each trapping potential.}
    \label{fig:trap}
\end{figure}

In this paper we generalize previous results on boundary nucleation, by applying the stochastic projected Gross--Pitaevskii (SPGPE) methodology to finite-temperature, quasi-2D analogues in which the atoms are confined in a circular optical `bucket' trap \footnote{A circular `bucket' is chosen to best avoid particularly rapid seeded nucleation at sharp corners, which we observed in our preliminary simulations.}, as shown in Figure \ref{fig:trap}. Rather than restrict our approach to a particular analogue, we investigate all three of the pseudo-spin-1/2, spin-1 and ferromagnetic analogues described above. In a simple circular bucket trap, boundary nucleation in the first two analogue systems is preferential, to the extent that reliably measuring the bulk nucleation rate would be difficult in a system of any experimentally realistic size. However, we show that incorporating a \textit{trench} near the edge of the trapping potential strongly suppresses boundary nucleation, allowing one to observe nucleation events in the bulk of a realistic-sized system; this constitutes a vital step towards realizing an analogue experiment that probes the bulk bubble nucleation rate in these systems. In a companion paper \cite{Jenkins:2025}, the same conclusion is reached for the pseudo-spin-1/2 system using the Truncated Wigner methodology.

The remainder of this paper is structured as follows. In Sections II--IV we present --- in turn for the pseudo-spin-1/2, ferromagnetic, and spin-1 analogues --- more details of the theoretical descriptions and parameters used in our simulations. Our simulation results use both pure `bucket' traps and `bucket' traps with the added `trench'. In each of these sections we conduct simulations using parameters suitable for realization of the analogue with a particular atomic species, respectively: potassium-39, rubidium-87 and sodium-23. Section V comprises our conclusions.

\section{Pseudo-spin-1/2 analogue: potassium-39}
\label{sec:k39}

The potassium-39 system we consider here is similar to the one described in Ref \cite{Jenkins:2023npg}.
It is based on potassium-39 atoms occupying two Zeeman levels condensed in a two-dimensional optical `bucket' trap.
Atomic collisions between atoms in the same level, or different levels, are described by three scattering parameters
$g_{\uparrow\uparrow}$, $g_{\downarrow\downarrow}$ and $g_{\uparrow\downarrow}$. 
In addition, a time-modulated  microwave field provides mixing between the atoms in each level, 
described by a Rabi frequency
$\Omega$ and a dimensionless modulation parameter $\lambda$. The time-averaged Hamiltonian for the two-component mean field $\psi$ is
\begin{align}
H&=\int\left\{-\frac{\hbar^2}{2m}\psi^\dagger\nabla^2\psi
+\frac12\sum_{i,j} g_{ij}|\psi_i|^2|\psi_j|^2+
(V_T-\mu)\psi^\dagger \psi\right.\nonumber\\
&\left.-\frac{\hbar\Omega}{2}\psi^\dagger\sigma_x \psi
+\frac{\hbar\Omega}2g'(\psi^\dagger\sigma_y\psi)^2
\right\}dxdy.
\end{align}
Here, the scattering parameters $g_{ij}$ determine the relative number density of the two components in the
ground state of the system, $n_\uparrow $ and $n_\downarrow $, and $g'=\lambda^2/4\sqrt{n_\uparrow n_\downarrow}$. 
Important physical parameters are the frequency scale 
$\omega_m=(g_{\uparrow\uparrow}+g_{\downarrow\downarrow}-2g_{\uparrow\downarrow})n/2\hbar$, healing length $\xi_m=(\hbar/m\omega_m)^{1/2}$ and temperature scale $T_m=\hbar \omega_m/k_{\text{B}}$, where $n=n_\uparrow +n_\downarrow $ is the total 
number density.

In the first configuration, the trapping potential $V_T(x,y)$ forms a circular `bucket' trap. Outside the trap, the potential drives the density to zero. There is a narrow transition region at the edge of the trap, which we can arrange to  have a width approximately equal to $\xi_m$. The bulk density inside the trap has a constant value $n$
in the initial state. In the second configuration, a drop in potential, or trench, on the inner side of the circular edge increases the condensate density there. Figure \ref{fig:trap} shows these configurations.

In  \cite{Jenkins:2023npg}, it was shown that the system can be described by an effective theory for a single
scalar field related to the relative phase $\varphi$ of the two components.
The canonically normalised field $\phi=v \varphi$, where
\begin{equation}
v^2=n_\uparrow n_\downarrow \xi_m^2\hbar\omega_m /n.
\end{equation}
The field equation has Klein-Gordon form
\begin{equation}
c_\varphi^{-2}\ddot\phi-\nabla^2\phi+\frac{dV}{d\phi}=0,\label{KGequation}
\end{equation}
where the sound speed $c_\varphi=\xi_m\omega_m(n_\uparrow n_\downarrow )^{1/2}\sqrt{2}/n$ and the potential
\begin{equation}
V(\phi)=\hbar\Omega \sqrt{n_\uparrow n_\downarrow }\left(-\cos\frac\phi{v}+\frac12(\lambda^2-1)\sin^2\frac\phi{v}\right).\label{pot}
\end{equation}
When $\lambda^2>1$, the potential has a local minimum, or false vacuum, at $\phi=v\pi$ and a global minimum, or true vacuum,
at $\phi=0$. The energy density difference between the vaccua is $\epsilon=2\hbar\Omega\sqrt{n_\uparrow n_\downarrow }$.

For the finite temperature simulations, we solve the simple-growth variant of the stochastic, projected Gross-Pitaevski equation (SPGPE) for
the condensate fields $\psi_i$~\cite{GardinerStochastic2002,GardinerStochastic2003,bradley_bose-einstein_2008,BradleyStochastic2014}. The SPGPE is
\begin{equation}
i\hbar\frac{\partial\psi_i}{\partial t}={\cal P}\left\{(1+i\gamma)\frac{\partial H}{\partial \overline\psi_i}+\eta_i\right\},
\end{equation}
where $H$ is the Hamiltonian, $i\gamma$ is a dissipation term and $\eta_m$ is a Gaussian stochastic noise term
with statistics
\begin{equation}
\langle\eta_i({\bf r},t)\eta_j^\dagger({\bf r}',t')\rangle=\frac{2\gamma k_BT}{\hbar\omega_m n}\delta_{ij}\delta({\bf r}-{\bf r}')\delta(t-t').
\end{equation}
The equation is solved in a two-dimensional periodic box containing the circular trap with functional form 
\begin{equation}
    V_{\text{trap}} = \dfrac{1}{2} \Big[2\mu - \mu\tanh \left(\frac{r-r_a}{\sigma} \right) + 3\mu\tanh \left( \frac{r-r_b}{\sigma} \right), \Big].
    \label{eq:trap}
\end{equation}
For the pure `bucket' trap we use $r_a = r_b = 59.6\mu$m and $\sigma = 2.64\mu$m, whereas for the `trench' trap we use $r_a = 55.6\mu$m, $r_b = 59.6\mu$m, and $\sigma = 0.264\mu$m.
Averages of observables over many runs represent thermal ensemble averages. We make the conceptual leap that snapshots taken from individual runs are representative of what could be observed in a real experiment~\cite{blakie_dynamics_2008}.

The system is initialized in the false vacuum state. This is achieved by equilibrating in the true vacuum state
and subsequently changing the sign of the Rabi frequency $\Omega$ with a piecewise linear ramp, switching the
true and false vacua~\cite{Billam:2021psh}. Simulations have been run with a range of timings in the initialisation process to ensure robustness of the procedure. Times in the final results are expressed relative to the end of the ramp.

\begin{figure}[ht]
    \centering    
    \includegraphics[width=0.95\linewidth]{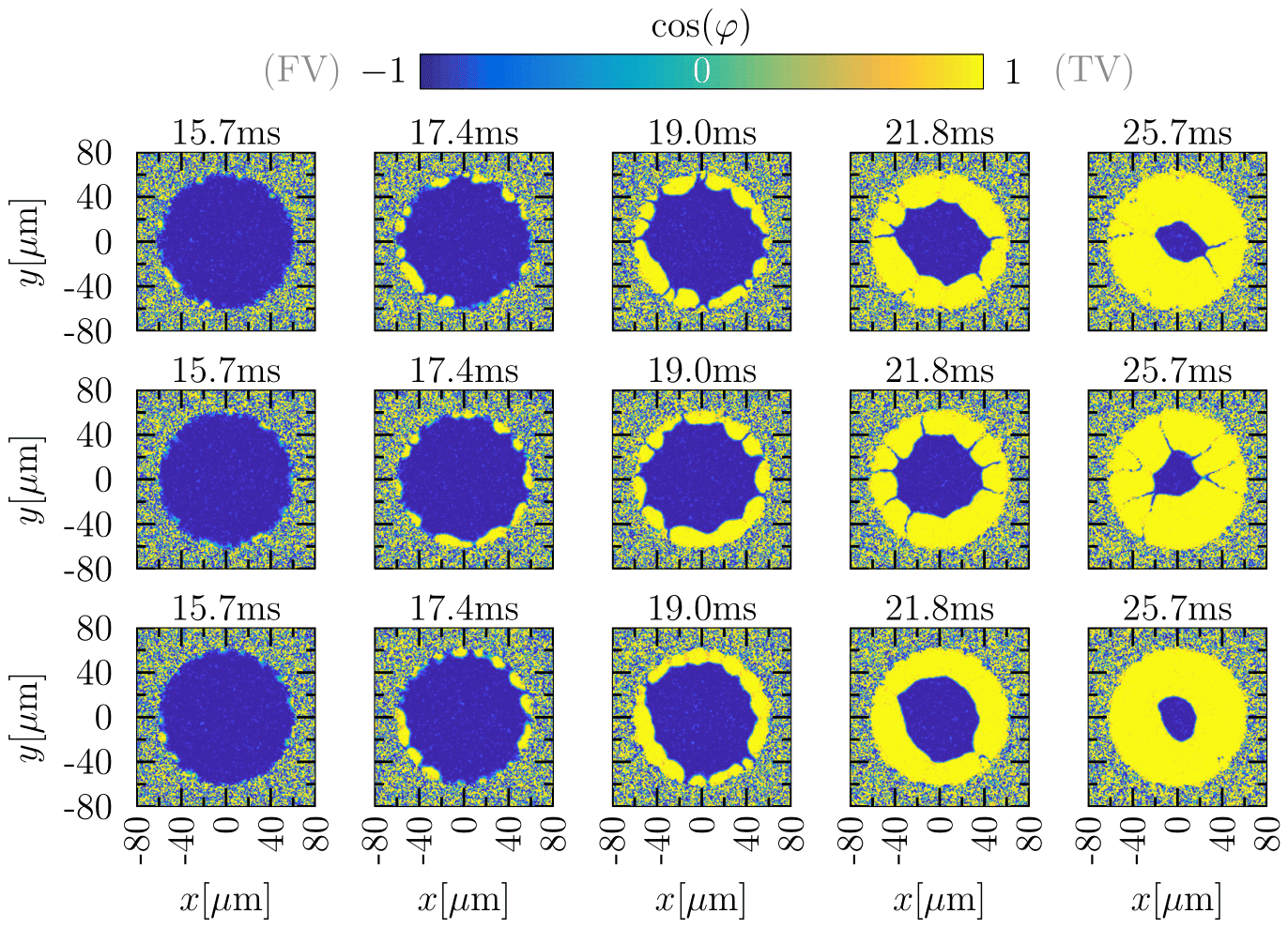}
    \caption{The evolution of $\cos(\varphi)$ in a pseudo-spin-1/2 gas of potassium-39 atoms confined to a circular `bucket’ potential at $T=48.50\mathrm{nK}$. Here $r_a = r_b = 59.6\mu\text{m}$ and $\sigma = 2.64\mu\text{m}$, with all relevant parameters listed in Table \ref{tab:K}. Each row of snapshots shows the progression of a unique simulation run, with time increasing from left to right. Identical parameters are used across realizations and all quantities are expressed in physical units.}
    \label{fig:K_bucket}
\end{figure}

A sequence of three runs at finite temperature in the pure `bucket' trap is shown in figure \ref{fig:K_bucket}, for parameters in table \ref{tab:K}. These show that the bubbles nucleate preferentially on the walls of the trap. This phenomenon can be understood in terms of nucleation theory, as mentioned in section~\ref{sec:intro}. There is a critical size for bubbles of the true vacuum phases, such that smaller bubbles collapse and larger bubbles grow. The bubble nucleation rate is exponentially suppressed by the energy of the critical bubble. However, near to a boundary, it is possible to have a fraction of a bubble with a fraction of the energy, which has far less suppression. The simulations show that this fraction is approximately one half, as predicted in a recent paper on seeded nucleation of bubbles \cite{Caneletti:2024kww}.

\begin{figure}[ht]
    \centering    
    \includegraphics[width=0.95\linewidth]{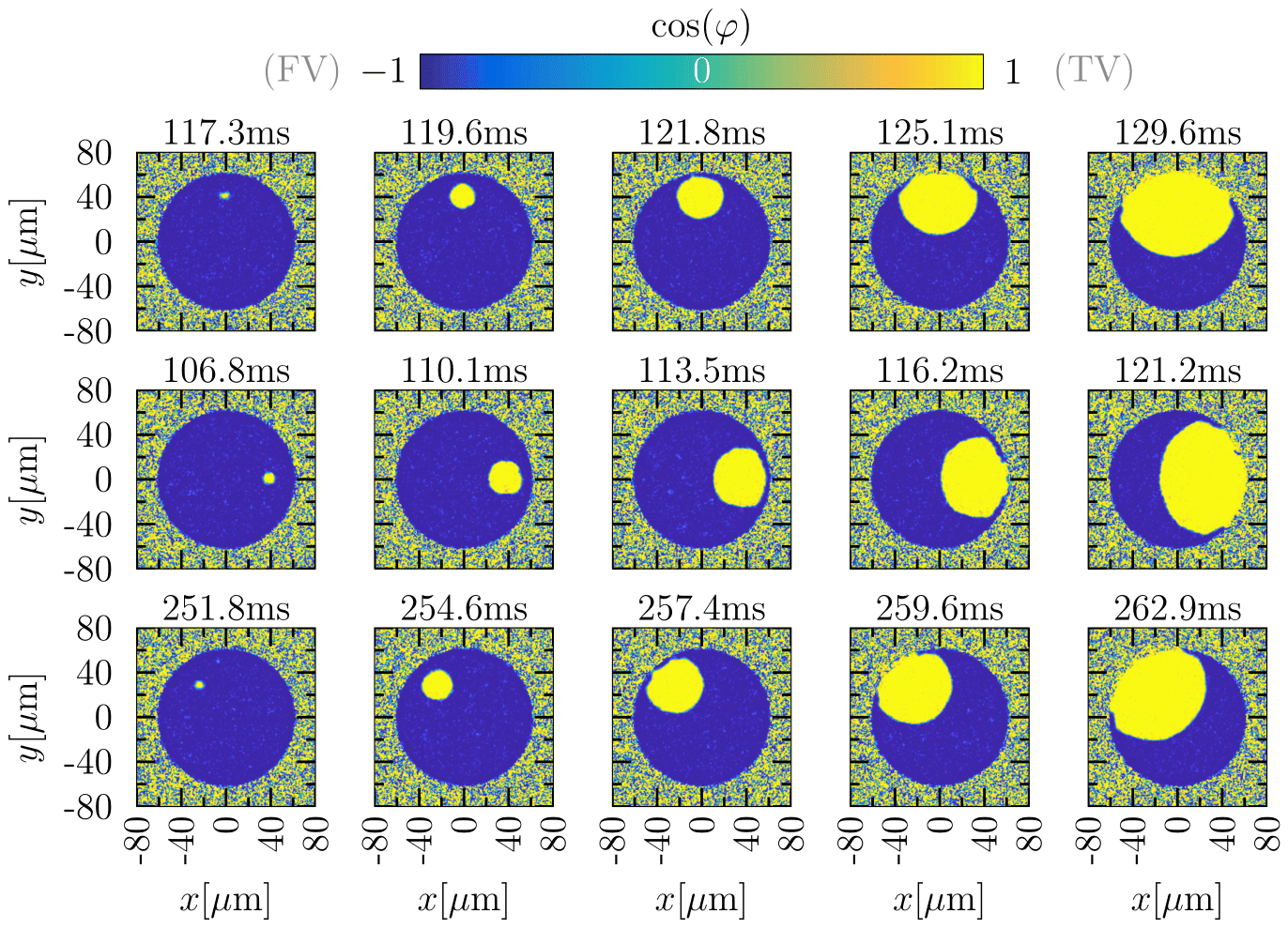}
    \caption{The evolution of $\cos(\varphi)$ in a pseudo-spin-1/2 gas of potassium-39 atoms confined to a`trench’ potential at $T=48.50\mathrm{nK}$. Here $r_a = 55.6\mu\text{m}$, $r_b = 59.6\mu\text{m}$, and $\sigma = 0.264\mu\text{m}$, with all relevant parameters listed in Table \ref{tab:K}. Each row of snapshots shows the progression of a unique simulation run, with time increasing from left to right. Identical parameters are used across realizations and all quantities are expressed in physical units.}
    \label{fig:K_trench}
\end{figure}

To observe bubbles nucleating inside the trap, rather than at the edges, we add a `trench' in the potential.
As described in section~\ref{sec:intro}, this screens the edge of the trap with a region of higher atomic density that acts like an anti-surfactant to prevent bubbles nucleating at the edges. The sequence of runs in figure \ref{fig:K_trench} shows the effect of having a trench in the potential. The potential has been chosen so that the density inside the trench is twice the density inside the rest of the trap. We found that changing the shape of the trench, whilst maintaining its qualitative features, did not affect our conclusions. 

\begin{table}
\caption{\label{tab:K39constants}Physical parameters used in the potassium-39 simulations.}
\begin{ruledtabular}\label{tab:K}.
\begin{tabular}{ll}
Parameter & Value \\
\hline  \\ [-1.5ex]
number density&$n=100\,\mu{\rm m}^{-2}$\\
magnetic field&$B=57.5{\rm G}$\\
Rabi frequency&$\Omega=100\times2\pi{\rm Hz}$\\
trap frequency&$\omega_\perp=5\times2\pi{\rm kHz}$\\
frequency scale&$\omega_m=23.14{\rm kHz}$\\
healing length&$\xi_m=0.264\mu{\rm m}$\\
temperature scale &$T_m=186.53{\rm nK}$\\
\end{tabular}
\end{ruledtabular}
\end{table}

\section{Spin-1 analogue: rubidium-87}
\label{sec:rb87}

\begin{figure}[htb]
    \centering
    \includegraphics[width=0.3\linewidth]{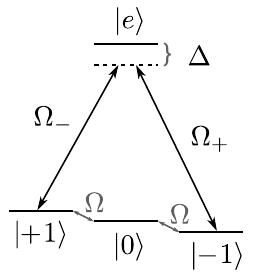}
    \caption{ Level coupling diagram for the spin-1 rubidium-87 system. The $F=1$ spin states, labeled $|m\rangle$, are coupled by a resonant RF beam with Rabi frequency $\Omega$, and by a two-photon Raman coupling induced by off-resonant optical beams with Rabi frequencies $\Omega_{\pm}$, zero two-photon detuning, and detuning $\Delta$ from the excited state $| e\rangle$}
    \label{fig:raman}
\end{figure}

The false vacuum decay proposal with rubidium-87 uses a three level system, with populations $n_+$, $n_0$ and $n_-$ in $m=+1$, $m=0$ and $m=-1$ Zeeman levels~\cite{Billam:2021nbc,Billam:2023}. The quadratic Zeeman effect, which plays an important role in determining the vacuum state of the system, is parameterised by a frequency $\omega_q$. The coupling between the levels is parameterised by just two parameters $g$ and $g'$, because of rotational symmetry. The microwave field mixing the levels is unmodulated, and an extra source of coupling between the levels is introduced using a Raman transition, parameterised by a constant $\lambda$, as shown in figure \ref{fig:raman}.
We define the natural frequency of the system to
be $\omega_m=gn$.

The Hamiltonian for the rubidium-87 system, when expressed in terms of a three component mean field $\psi$, is
\begin{equation}
H=\int\left\{-\frac{\hbar^2}{2m}\psi^\dagger\nabla^2\psi
+(V_T-\mu)\psi^\dagger \psi
+{\cal V}
\right\}dxdy.
\end{equation}
where the scattering and interaction terms are
\begin{align}
{\cal V}&=\hbar\omega_q(\overline\psi J_z^2\psi)
+\frac12g(\overline\psi\psi)^2+\frac12g'(\overline\psi {\bf J}\,\psi)^2\notag\\
&+\frac12\hbar\Omega\overline\psi J_x\psi+
-\frac18\hbar\lambda^2\overline\psi \left(J_+^2+J_-^2\right)\psi.
\end{align}
The 2D scattering coefficients are related to scattering lengths in the $F$ channels $a_0$ and $a_2$ \cite{Kawaguchi2012,Stamper-Kurn2013}
\begin{equation}
g=\left(\frac{8\pi\hbar^3\omega_\perp}{m}\right)^{1/2}\frac{a_0+2a_2}{3},\quad
g'=\left(\frac{8\pi\hbar^3\omega_\perp}{m}\right)^{1/2}\frac{a_2-a_0}{3}.
\end{equation}
For rubidium-87, the ratio $g'/g=-0.00463$ . The quadratic Zeeman energy
\begin{equation}
\hbar\omega_q=\frac{(g_F\mu_BB)^2}{\Delta E_{\rm hfs}},
\end{equation}
where $g_F\approx1/2$ and $\mu_B$ is the Bohr magneton. The Raman coupling coefficient $\lambda$
was shown to be
\begin{equation}
\lambda^2=\frac{\Omega_+\Omega_+}
{\Omega\Delta_e}.
\end{equation}
where $\Omega$ is the RF Rabi frequency, $\Omega_\pm$ the optical Rabi frequency and $\Delta_e$
the detuning. 

The state of lowest energy when $g'<0$, $g>0$ and 
$0<\hbar\omega_q<-2g'n$ is called the broken axisymmetric (BA) phase, in which all
three Zeeman levels are occupied. There is an effective theory for the relative phase $\varphi$ of the $m=+1$
and $m=-1$ components of the BEC in the BA phase. The field equation has Klein-Gordon form (\ref{KGequation})
with potential
\begin{equation}
V=2 \hbar\Omega \,n_+\left(\lambda_c^2\cos\varphi+\frac12\lambda^2\sin^2\varphi\right),
\end{equation}
where
\begin{equation}
\lambda_c=\left(\frac{1-\hbar\omega_q/2ng'}
{1+\hbar\omega_q/2ng'}\right)^{1/2}.
\end{equation}

\begin{figure}[ht]
    \centering
    \includegraphics[width=0.95\linewidth]{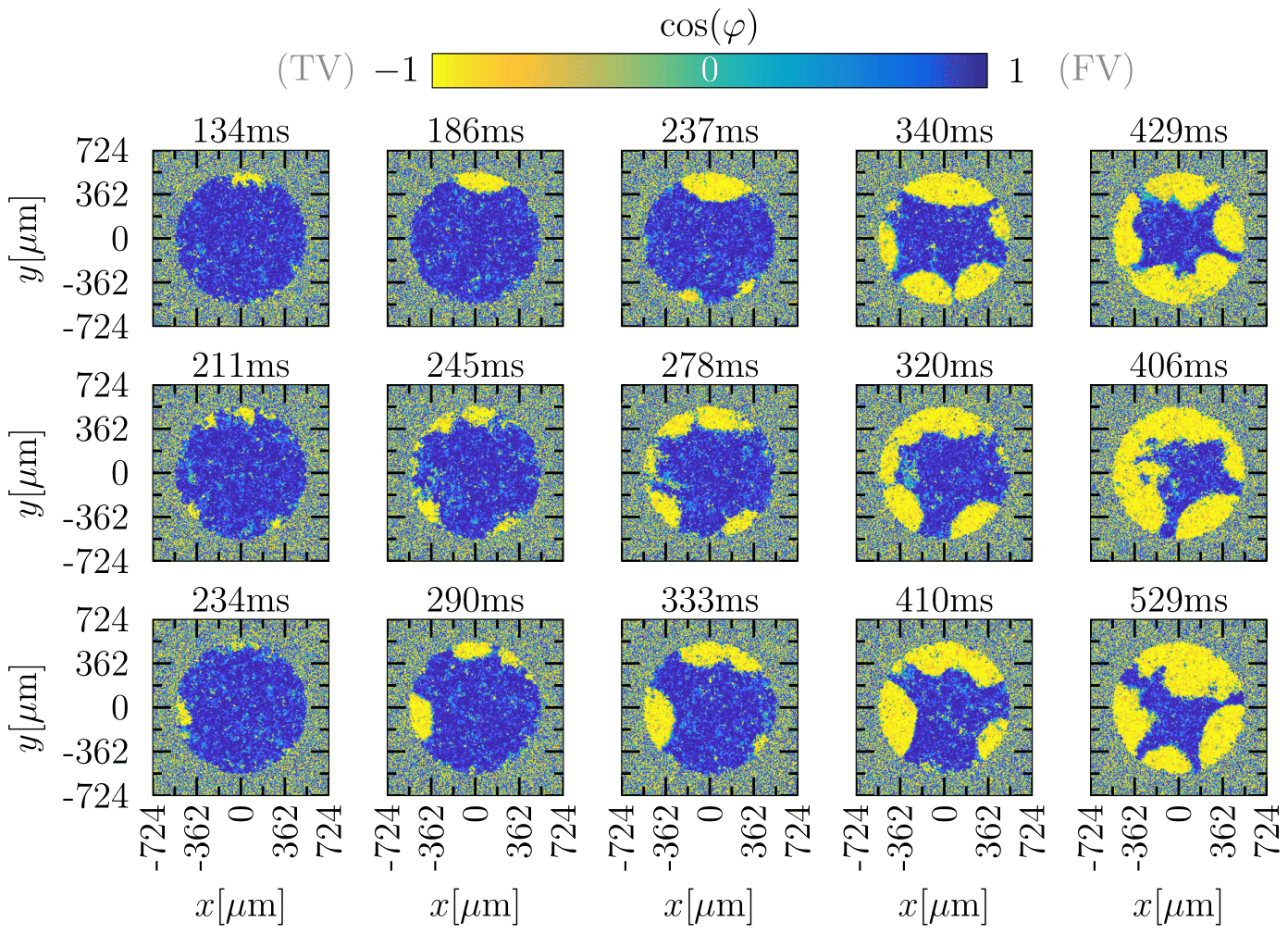}    
    \caption{The evolution of $\cos(\varphi)$ in a spin-1 gas of rubidium-87 atoms confined to a circular `bucket’ potential at $T=16.56\mathrm{nK}$. Here $r_a = r_b = 543\mu\text{m}$ and $\sigma = 18.1\mu\text{m}$, with all relevant parameters listed in Table \ref{tab:Rb}. Each row of snapshots shows the progression of a unique simulation run, with time increasing from left to right. Identical parameters are used across realizations and all quantities are expressed in physical units.}
    \label{fig:Rb_bucket}
\end{figure}

\begin{figure}[ht]
    \centering
    \includegraphics[width=0.95\linewidth]{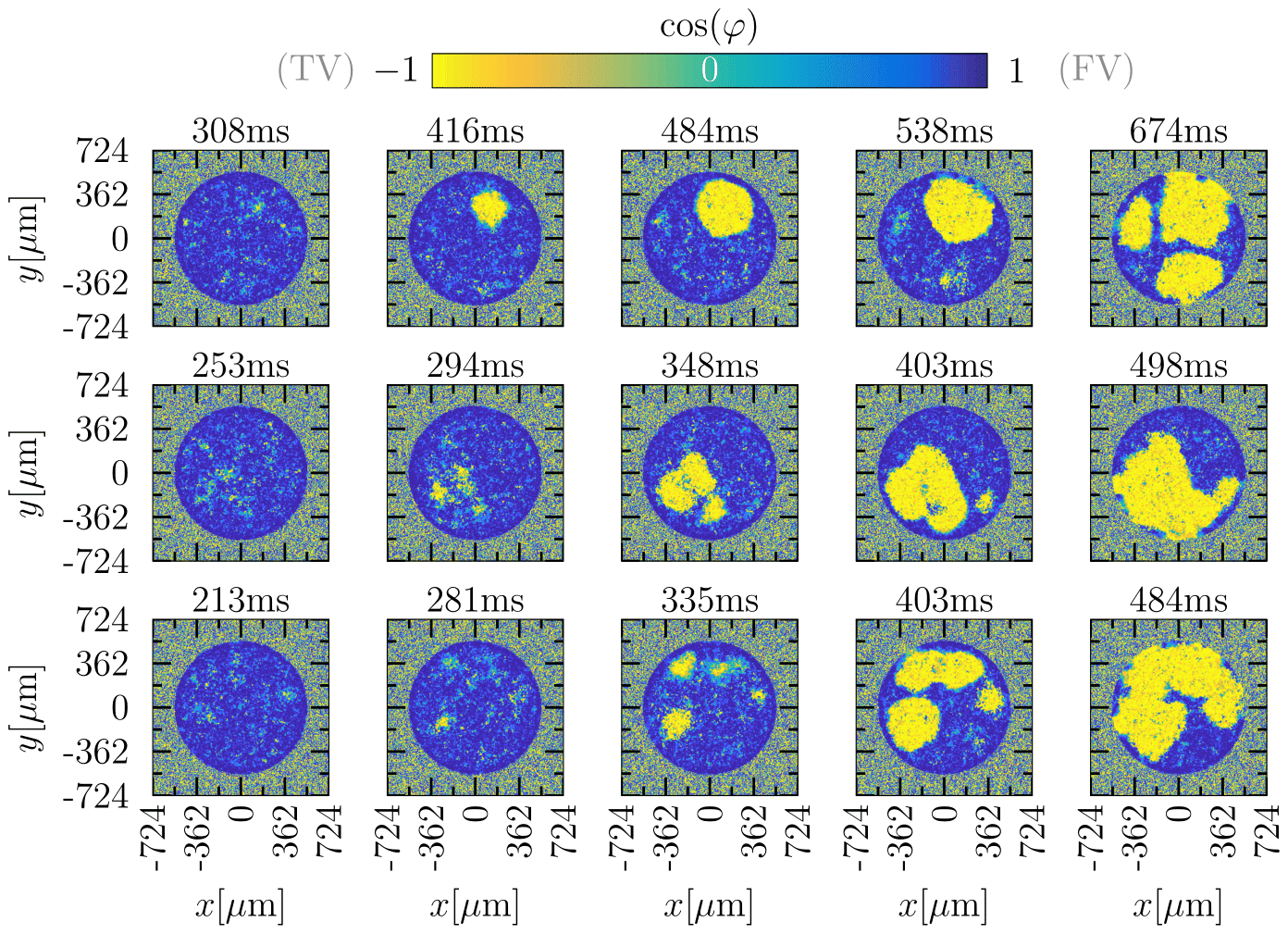}
    \caption{The evolution of $\cos(\varphi)$ in a spin-1 gas of rubidium-87 atoms confined to a `trench’ potential at $T=16.56\mathrm{nK}$. Here $r_a = 507\mu\text{m}$, $r_b = 543\mu\text{m}$, and $\sigma = 1.81\mu\text{m}$, with all relevant parameters listed in Table \ref{tab:Rb}. Each row of snapshots shows the progression of a unique simulation run, with time increasing from left to right. Identical parameters are used across realizations and all quantities are expressed in physical units.}
    \label{fig:Rb_trench}
\end{figure}

A sequence of three runs for the `bucket' trap potential is shown in figure \ref{fig:Rb_bucket}, for parameters in table \ref{tab:Rb}. Again, the length parameters in the potential are chosen the same relative to the healing length as they were in section~\ref{sec:k39}. These sequences show that the bubbles nucleate preferentially on the walls of the trap. The sequence of runs in figure \ref{fig:Rb_trench} shows the effect of having the `trench' in the potential, which again acts to suppress nucleation at the boundaries.

\begin{table}[ht]
\caption{Physical parameters used in the rubidium-87 simulations.}
\begin{ruledtabular}
\begin{tabular}{ll}\label{tab:Rb}
Parameter & Value \\
\hline  \\ [-1.5ex]
number density&$n=100\,\mu{\rm m}^{-2}$\\
magnetic field&$B=0.389{\rm G}$\\
Rabi frequency&$\Omega=5\times2\pi{\rm Hz}$\\
Raman parameter&$\lambda=1.45$\\
trap frequency&$\omega_\perp=15\times2\pi{\rm kHz}$\\
frequency scale&$\omega_m=22.1{\rm kHz}$\\
healing length&$\xi_m=1.81\mu{\rm m}$\\
temperature scale&$T_m=55.2{\rm nK}$\\
\end{tabular}
\end{ruledtabular}
\end{table}

The small value of the ratio $g'/g$ results in parameter choices for the rubidium-87 system that are especially challenging. Parameter values for related systems using potassium-41 or lithium-7 have the potential to be more experimentally favorable, but this has to be offset by the ease of cooling rubidium-87 compared to these alternative species.

\section{Ferromagnetic analogue: sodium-23}

Sodium-23 is the only BEC system so far with actual experimental results on bubble formation~\cite{Zenesini:2024}.
We will model a version of the system which is extended into two dimensions.
The basic system has two hyperfine levels. In this case the
frequency $\omega_m$ is negative. The microwave field mixing the levels is unmodulated, but 
is detuned by an amount $\delta$ from the frequency separation between the two levels. The 
mean field Hamiltonian for the two-component mean field $\psi$ is
\begin{align}
H=&\int\left\{-\frac{\hbar^2}{2m}\psi^\dagger\nabla^2\psi
+\frac12\sum_{i,j} g_{ij}|\psi_i|^2|\psi_j|^2+
(V_T-\mu)\psi^\dagger \psi\right.\nonumber\\
&\left.-\frac{\hbar\Omega}{2}\psi^\dagger\sigma_x \psi
+\frac{\hbar\delta}{2}\psi^\dagger\sigma_z \psi
\right\}dxdy.
\end{align}

The effective theory is now in the magnetisation sector $Z$, where
\begin{equation}
Z=\frac{n_\uparrow -n_\downarrow }{n_\uparrow +n_\downarrow }.
\end{equation}
The canonically normalised field is $\phi=(\hbar^2 n/4m)^{1/2}\cos^{-1}(Z)$, with field equation
\begin{equation}
    c_z^{-1}\left(c_z^{-1}\dot\phi\right)\dot{}-\nabla^2\phi+{\frac{\partial V}{\partial \phi}}.
\end{equation}
The potential is
\begin{equation}
    V(Z)=\frac{n}{4}\left[\hbar\omega_m Z^2-2\hbar\Omega\sqrt{1-Z^2}-2\hbar(\delta+\omega_m) \,Z
    \right].\label{NaV}
\end{equation}
There is a critical value for the detuning $\delta_c$ at which the potential develops a false vacuum minimum.

In this model, the sound speed is dependent on $Z$, 
\begin{equation}
    c_z=\sqrt{\frac{\hbar\Omega}{2m}}(1-Z^2)^{-1/4}.
\end{equation}
This field dependence in the sound speed breaks the Lorentz symmetry. This is not a significant issue for finite temperature nucleation, where the instanton is independent of imaginary time and the value of the sound speed $c_z$ is immaterial. 

At zero temperature, it is undesirable for an analogue system describing elementary particle physics to break Lorentz invariance. However, the system has features that make it resemble
one with Lorentz symmetry. Firstly, the nucleation of bubbles is very close to the Lorentz invariant case if the barrier is very narrow, and $Z$ remains close to the false vacuum value inside the bubble instanton. Secondly, bubbles grow in a very similar way to the bubbles in a Lorentz invariant theory (shown in appendix \ref{apA}).

\begin{figure}[htb]
    \centering
    \includegraphics[width=0.95\linewidth]{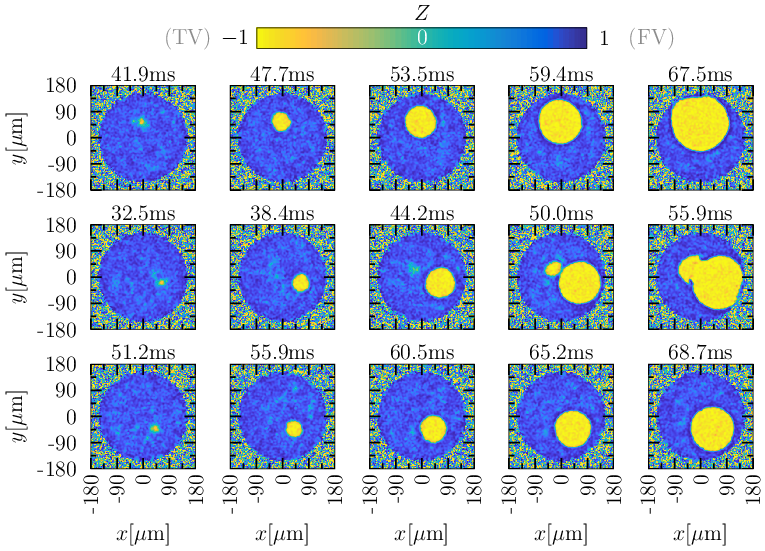}
    \caption{The evolution of the magnetization, $Z$, in a pseudo-spin-1/2 gas of sodium-23 atoms confined to a circular `bucket’ potential at $T=5\mathrm{nK}$. Here $r_a = r_b =150\mu\text{m}$ and $\sigma = 15\mu\text{m}$, with all relevant parameters listed in Table \ref{tab:Na}. Each row of snapshots shows the progression of a unique simulation run, with time increasing from left to right. Identical parameters are used across realizations and all quantities are expressed in physical units.}
    \label{fig:Na_bucket}
\end{figure}

A sequence of three runs for the `bucket' trap is shown in figure \ref{fig:Na_bucket}, for parameters in table \ref{tab:Na}. The length parameters in the potential ($r_a$, $r_b$, and $\sigma$) are chosen to be the same relative to the healing length as they were in section~\ref{sec:k39}.  These show that the bubbles nucleate preferentially inside the trap rather than at the walls. The reason for this can be traced back to the potential (\ref{NaV}), in which the frequency $\omega_m$ depends on density. At low density near the edge of the trap, the false vacuum becomes a true vacuum and disfavors bubble nucleation there.

The runs also show an interesting feature of the sodium-23 system that vortices can form when the bubble begins to self-intersect In larger systems we expect vortices to form when the bubbles collide. This interesting phenomenon offers new insight into the possibility of creating topological structures in the very early universe. 

It is also apparent that the bubbles in the sodium-23 system become distorted quite early on in their growth phase. This is due to the bubble wall becoming thinner as the bubble expands, and the large spatial gradients in the wall cause the breakdown of the effective theory. This phenomenon is not restricted to the sodium-23 system, and happens eventually to bubbles in all the analogue vacuum decay systems. In each system there is a small parameter that controls the validity of the effective theory. In the potassium-39 system of section~\ref{sec:k39} and the rubidium-87 system of section~\ref{sec:rb87}) the parameter is $\Omega/\omega_m$, which can be tuned to be as small as we like, and was 0.008 in the potassium-39 simulations. In the sodium-23 system, the parameter is fixed by the scattering lengths. If we define 
$\omega_n=(g_{\uparrow\uparrow}+g_{\downarrow\downarrow}+2g_{\uparrow\downarrow})n/\hbar$,
then the small parameter is $|\omega_m/\omega_n|=0.04$.

\begin{table}[htb]
\caption{\label{tab:Na23constants}Physical parameters used in the sodium-23 simulations.}
\begin{ruledtabular}
\begin{tabular}{ll}\label{tab:Na}
Parameter & Value \\
\hline  \\ [-1.5ex]
number density&$n=100\,\mu{\rm m}^{-2}$\\
Rabi frequency&$\Omega=100\times2\pi{\rm Hz}$\\
detuning&$\delta-\delta_c=68\times2\pi{\rm Hz}$\\
trap frequency&$\omega_\perp=5.0\times2\pi{\rm kHz}$\\
frequency scale&$\omega_m=-1.2{\rm kHz}$\\
healing length&$\xi_m=1.51\mu{\rm m}$\\
temperature scale&$T_m=9.7{\rm nK}$\\
\end{tabular}
\end{ruledtabular}
\end{table}

\section{Conclusions}

In order to gain insight into the physics of false vacuum decay in the early universe from a cold-atom analogue simulator, a key next step is to reliably probe experimentally the bulk bubble nucleation rate in a homogeneous condensate. In this paper we described three previously proposed analogue systems, and found issues with rapid bubble nucleation at the trap boundary for pseudo-spin-1/2 (modelled here for potassium-39) and spin-1 (modelled here for rubidium-87) analogues. For the case of an optically-trapped system at finite temperature our SPGPE simulations show that, in these analogues, this deleterious boundary nucleation can be mitigated by adding a `trench' to the potential, effectively screening the boundary with a region of higher atomic density than found in the uniform bulk of the system. Our results suggest that a potential with such a `trench', which can be created using digital micromirror devices~\cite{Gauthier2016} --may allow forthcoming experiments to measure bulk bubble nucleation rates in realistically sized systems.

Data supporting this publication are openly available under
a Creative Commons CC-BY-4.0 License in \cite{data_package_arxiv}.

\section*{Acknowledgements}
This work was supported by the Science and Technology Facilities Council (STFC) [Grant
ST/T000708/1] and the UK Quantum Technologies for Fundamental Physics programme [Grants
ST/T00584X/1 and ST/W006162/1]. The authors are grateful to Russell Bisset for discussions on sculpting the potential, and to Gabriele Ferrari, Alex Jenkins, Hiranya Peiris, Andrew Pontzen and Zoran Hadzibabic for insightful conversations. This research made use of the Rocket High Performance Computing service at Newcastle University.

\appendix

\section{Bubble wall motion in pseudo-Lorentzian theories}\label{apA}

The sodium-23 system is an example where the effective theory is not Lorentz invariant, but bubble walls expand in a way that resembles relativistic analogue systems. We will investigate the motion of bubble walls in such theories by constructing an action for the bubble radius.

The family of theories of interest has Lagrangian density ${\cal L}$ of the form
\begin{equation}
    {\cal L}=\frac12F(Z)\,\dot Z^2-\frac12G(Z)\,(\nabla Z)^2-V(Z)+V(Z_{FV})
\end{equation}
Lorentz invariance applies when $G(Z)=c^2F(Z)$, where $c$ is the constant sound speed. Consider the more general case where this condition is not imposed. We apply a thin-wall anzatz for the field that extrapolates between the true vacuum value $Z_{TV}$ and the false vacuum value $Z_{FV}$
\begin{equation}
    Z=\frac12Z_{TV}(1-f)+\frac12Z_{FV}(1+f)
\end{equation}
where $f$ depends on the bubble radius $R(t)$, a scaling for the wall thickness $\gamma(t)$ and an overall scaling constant $\mu$,
\begin{equation}
    f=\tanh\left[\mu\gamma(t)(r-R(t))\right]
\end{equation}
This approach depends on the fact that the action is stationary, and can be approximated with a relatively crude approximation to the field. Due to the thin-wall approximation, we will drop terms with $(1-f^2)(r-R)$ factors.
We have
\begin{align*}
\dot Z&\approx -\frac{\mu\gamma}{2}(1-f^2)\Delta Z\,\dot R\\
Z'&\approx \frac{\mu\gamma}{2}(1-f^2)\Delta Z\\
\end{align*}
where $\Delta Z=Z_{FV}-Z_{TV}$. We also assume a similar anzatz for the potential $V(Z)$, but with and extra term representing the potential barrier around $r=R$, and so we write
\begin{equation}
    V=\frac12V_{TV}(1-f)+\frac12V_{FV}(1+f)+\Delta V(x)
\end{equation}
where $x=\mu\gamma(r-R)$.

We define the Lagrangian $L$ in $n$ spatial dimensions by
\begin{equation}
    L=\int{\cal L}\,d^nx
\end{equation}
After inserting the anzatzes, we arrive at
\begin{align}
    L=&\frac{\omega_n}{\mu\gamma}R^{n-1}\left\{
    \frac12\left(\frac{\mu\gamma}{2}\right)^2(\Delta Z)^2c_F\dot R^2
    -\frac12\left(\frac{\mu\gamma}{2}\right)^2(\Delta Z)^2c_G -V_B\right\}
    \nonumber\\
    &-\frac{\omega_n}{n}R^n\epsilon
\end{align}
where $\omega_n$ is the area of a sphere in $n$ dimensions, $\epsilon=V_{FV}-V_{TV}$, and
\begin{align}
c_F&=\int F(Z)(1-f^2)^2 dx\\
c_G&=\int G(Z)(1-f^2)^2 dx\\
V_B&=\int \Delta V \,dx
\end{align}
Now we make a judicious choice of $\mu$ to arrange that
\begin{equation}
    L=\frac{\omega_n}{2}\sigma\gamma\left(\dot R^2/c^2-1-\gamma^{-2}\right)R^{n-1}+
    \frac{\omega_n}{n}\epsilon R^n.
\end{equation}
In this formula, $c=(c_G/c_F)^{1/2}$ and
\begin{align}
\mu&=\sqrt{\frac{8V_B}{c_G\Delta Z^2}}\\
\sigma&=\sqrt{\frac{c_GV_B\Delta Z^2}{2}}
\end{align}
The action should be stationary with respect to variations of $\gamma$, which implies
\begin{equation}
\gamma=(1-\dot R^2/c^2)^{-1/2}
\end{equation}
We find that the thickness of the bubble is Lorentz contracted, just as in a relativistic theory.
Substituting back into the action,
\begin{equation}
    L=-\omega_n\sigma(1-\dot R^2/c^2)^{-1/2}R^{n-1}+
    \frac{\omega_n}{n}\epsilon R^n
\end{equation}
The effective theory of the bubble wall is the same as the one we would obtain from a fully Lorentz invariant theory, with propagation speed $c$ and surface tension $\sigma$. There is a caveat, however, in that all the analogue systems have terms of the form $\dot Z^2\nabla Z^2$ which can be neglected initially but eventualy break the Lorentz invariance when $\gamma$ becomes large.


%

\end{document}